\documentclass[12pt]{iopart}
\usepackage[T1]{fontenc}
\usepackage{graphicx}
\usepackage{mathptmx}
\begin{document}

\title[Transverse Momentum and Tsallis Distribution]{Transverse Momentum Distributions in proton - proton Collisions at LHC Energies
and Tsallis Thermodynamics}
\author{M.D. Azmi$^1$ and J. Cleymans$^2$}
\address{$^1$HEP Lab, Department of Physics, Aligarh Muslim University, Aligarh - 202002, India}
\address{$^2$ UCT-CERN Research Centre and Department of Physics,
University of Cape Town, Rondebosch 7701, South Africa}

\begin{abstract}
\noindent A detailed study of the transverse momentum distributions of charged particles produced in 
proton - proton 
collisions at LHC energies is presented. This is done using a thermodynamically consistent form of the Tsallis
distribution. 
All variables used  are  
thermodynamical and 
in particular, the temperature, $T$, follows from the standard thermodynamic definition as being  the derivative of the 
energy with respect to the (Tsallis) entropy.  The momentum distribution of the final state particles can be 
described very well by the Tsallis distribution. 
The values of the parameters are determined from measurements by  the ALICE, ATLAS and CMS 
collaborations and are discussed in detail. In particular, the Tsallis parameter, $q$, is found with consistent values  
for all the transverse momentum distributions despite large differences in kinematic regions and shows a slight
increase with  beam energy, reaching a value of 1.15 at 7 TeV.     
It is concluded that the hadronic system created in high-energy p - p collisions at mid-rapidity
can be seen as obeying Tsallis 
thermodynamics.
\end{abstract}
\maketitle
\section{Introduction}
The dynamics of colliding hadrons can be determined from the distribution of the momenta of 
produced particles.   
The available range of transverse momenta has expanded 
considerably with the advent of 
the Large Hadron Collider (LHC) at CERN. 
Collider energies up to 7 TeV are now available in proton - proton collisions and transverse momenta of 
hundreds of GeV are of a common occurrence. This  helps in testing and applying the relativistic 
thermodynamics and hydrodynamics in a new energy region. 
Thermal models are part of the standard set of tools to analyze high-energy collisions and in this paper 
we investigate a thermodynamically consistent model based on the Tsallis distribution~\cite{tsallis1,biro}.
This leads to a 
power law distribution which is well suited to describe the transverse momenta measured in p - p collisions. 
The Tsallis 
distribution was  first proposed about twenty-five years ago as a generalization of 
the usual exponential Boltzmann-Gibbs distribution, and is characterized by 
three parameters $q$, $T$ and $V$. 
\noindent Various versions of the Tsallis distribution have been considered in the 
literature~\cite{tsallis2,buyukilic,penini,plastino,conroy2,conroy1,TsallisBE}.
A form which is suited for describing results in  
high energy particle physics is used in this paper.  \\
\indent The variable $T$ used here obeys the standard thermodynamic relation~\cite{worku1,worku2},
\begin{equation}
	\label{defineT}
	T = \left.\frac{\partial E}{\partial S}\right|_{N,V},
\end{equation}
and can, therefore, be referred as temperature. However, since the entropy used  in 
Eq.~(\ref{defineT})
is the Tsallis entropy~\cite{tsallis1} and not the standard entropy, we will call the variable 
defined in Eq.~(\ref{defineT}) as q-temperature.  
For a similar reason the volume $V$ can be obtained from thermodynamic relations as, 
\begin{equation}
	\label{defineV}
	V = \left.\frac{\partial H}{\partial P}\right|_{N,S},
\end{equation}
with $H = E + PV$ is the standard definition of the  enthalpy and 
	 could  be called q-volume as it is obtained at fixed Tsallis entropy but  
we  refrain from doing so and refer to it as volume but note that it is not necessarily related to a volume 
deduced from other models, say HBT calculations.\\
\indent A lot of interest has developed recently~\cite{biro1,wilk1,wilk2,deppman,dubna} in 
the Tsallis distribution
and it has been  
used previously to describe the transverse momentum data of the charged particles produced in proton - proton 
collisions at RHIC and LHC energies~\cite{STAR,PHENIX1,PHENIX2,ALICE,CMS1,CMS2,ATLAS,ALICE2}. 
A comparative study, which has not been  done before, of the Tsallis fits to the transverse momentum 
distributions 
of the charged particles produced in proton - proton collisions measured by ALICE, ATLAS and CMS 
collaborations is presented here. The parameters used in the Tsallis fit are studied at different 
energies and under different kinematical conditions of the data collection. The parameters are measured at 
various multiplicities. The fit results the values of Tsallis parameter, $q$, in the range $1.1$ to $1.15$ for 
all the measured distributions and found to be  consistent in all the conditions and at all energies. 
The $T$ and $R$ parameters show dependence on the multiplicity and on particle yields too.
The results presented here confirm that the hadronic system created in high-energy p - p collisions obeys Tsallis thermodynamics.
\section{Tsallis Distribution}
\indent The transverse momentum distribution in heavy-ion collisions is often described by a combination of 
transverse flow and a thermodynamical statistical distribution. 
In p - p collisions with the Tsallis distribution such a  superposition is not needed
and very  good fits can be obtained. 

\indent In the framework of Tsallis statistics~\cite{tsallis1,conroy1,biro1,worku1, worku2}
integrals over 
\begin{equation}
f = \left[1 + (q-1)\frac{E-\mu}{T}\right]^{-\frac{1}{q-1}} 
\label{realtsallis}
\end{equation}
give the entropy, $S$, the particle number, $N$, the energy density, $\epsilon$, and the pressure, $P$.\\
\indent Using the function 
$$
\ln_{q}(x)\equiv\frac{x^{1-q} - 1}{1-q},
$$
often referred to as q-logarithm, it can be shown\cite{worku2} that the relevant 
thermodynamic quantities are given by:
\begin{eqnarray}
S &=& - gV\int\frac{d^3p}{(2\pi)^3}\left[f^{q}{\rm ln}_{q}f - f\right],\label{entropy}\\
N &=& gV\int\frac{d^3p}{(2\pi)^3} f^{q},\label{Number}\\
\epsilon &=& g\int\frac{d^3p}{(2\pi)^3}~E~f^{q},\label{epsilon}\\
P &=& g\int\frac{d^3p}{(2\pi)^3}\frac{p^{2}}{3E}~f^{q}.\label{pressure}
\end{eqnarray}
where $V$ is the volume and  $g$ is the degeneracy factor. \\
\indent In order to use the above equations it has to be shown that they satisfy the thermodynamic 
consistency conditions. The first and second laws of thermodynamics lead to the following two 
differential relations~\cite{groot}:
\begin{eqnarray}
d\epsilon=T~ds +\mu~dn,\label{eq1}\\
dP=s~dT + n~d\mu.\label{eq2}
\end{eqnarray}
where, $s = S/V$ and $n = N/V$ are the entropy and particle number 
densities, respectively.\\
\indent Thermodynamic consistency requires that the following relations be satisfied:
\begin{eqnarray}
T=\left.\frac{\partial\epsilon}{\partial s}\right|_n,\label{temp}\\
\mu=\left.\frac{\partial\epsilon}{\partial n}\right|_s,\label{mu}\\
n=\left.\frac{\partial P}{\partial \mu}\right|_T,\label{number}\\
s=\left.\frac{\partial P}{\partial T}\right|_\mu.\label{ent}
\end{eqnarray}
\indent Eq.~(\ref{temp}) in particular shows that the variable $T$ appearing in Eq.~(\ref{realtsallis}) indeed
can be identified as a thermodynamic temperature. As explained in the introduction we prefer 
to call it q-temperature as it is based on the Tsallis form of the entropy~\cite{tsallis1}.
It is straightforward to show that these relations are indeed satisfied~\cite{worku2}.

It can easily be shown that the following thermodynamic consistency relation is also satisfied:
\begin{equation}
\epsilon + P = T s + \mu n  .
\end{equation}
\indent The momentum distribution obtained from Eq.~(\ref{Number}) is given by:
\begin{equation}
\frac{d^3N}{d^3p}=\frac{gV}{(2\pi)^3}\left[1 + (q-1)\frac{E-\mu}{T}\right]^{-q/(q-1)},
\end{equation}
or expressed in terms of variables used in high-energy physics, 
transverse momentum, $p_T$, transverse mass, $m_T = \sqrt{{p_T}^2 + m^2}$ and rapidity, $y$:
\begin{equation}
\frac{d^2 N}{dp_T dy}=gV\frac{p_T~m_T \cosh y}{(2\pi)^2}
\left[1 + (q-1)\frac{m_T{\rm cosh}y - \mu}{T}\right]^{-q/(q-1)}
\label{ptdist}
\end{equation}
\indent At mid-rapidity, $y$ = 0, and for zero chemical potential, as is relevant at the LHC, Eq.~(\ref{ptdist})
reduces to the following expression:
\begin{eqnarray}
\left.\frac{d^{2}N}{dp_T~dy}\right|_{y=0} = gV\frac{p_Tm_T}{(2\pi)^2}\left[1+(q-1)\frac{m_T}{T}\right]^{-q/(q-1)}
\label{tsallisfit}
\end{eqnarray}
\indent It is worth to mention that the parameterization given in Eq.~(\ref{tsallisfit}) is close to 
the parameterization 
used for fitting the data taken at RHIC and LHC 
experiments~\cite{STAR,PHENIX1,PHENIX2,ALICE,CMS1,CMS2,ATLAS,ALICE2}. 
The parameterization used by the RHIC and LHC experiments is given below: 
\begin{equation}
\frac{d^2 N}{dp_T dy}=p_T\frac{dN}{dy}\frac{(n-1)(n-2)}{nC(nC + m_{0}(n-2))}
\left[1 + \frac{m_T - m_0}{nC}\right]^{-n},
\label{BIG}
\end{equation}
where $n, C$ and $m_0$ are the fit parameters used in the parameterization. \\
\indent At mid-rapidity, $y = 0$, and for zero chemical potential Eq.~(\ref{BIG}) shows 
the same dependence on the 
transverse momentum as Eq.~(\ref{tsallisfit}) except for an additional factor $m_T$ which is present in 
Eq.~(\ref{tsallisfit}) but not in Eq.~(\ref{BIG}). 

\section{Fit details}
The transverse momentum distributions of charged particles produced in $p - p$ collisions at LHC 
energies were fitted using a sum of three Tsallis distributions. 
These consist of fits for $\pi^+$'s, $K^+$'s and protons, $p$. 
The following expression, at mid-rapidity and $\mu = 0$, was used to fit the distributions 
obtained in various experiments:
\begin{equation}
\left.\frac{d^{2}N_{ch}}{dp_T~dy}\right|_{y=0} = 2p_T\frac{V}{(2\pi)^2}
\sum_{i=1}^{3}g_im_{T,i}\left[1+(q-1)\frac{m_{T,i}}{T}\right]^{\frac{-q}{(q-1)}}
\end{equation}
where $i = \pi^+, K^+, p$. The relative weights between particles were determined by the 
corresponding degeneracy factors and given by $g_{\pi^+}$ = $g_{K^+}$ = 1 and $g_p$ = 2. 
The factor 2 on the right hand side takes into account the contributions from 
antiparticles, $\pi^-, K^-$ and $\bar{p}$. \\
\indent In some cases the  following form of the distribution was used to fit the 
experimental data sets:
\begin{equation}
\left.\frac{1}{2\pi p_T}\frac{d^{2}N_{ch}}{dp_T~dy}\right|_{y=0} = \frac{2V}{(2\pi)^3}\sum_{i=1}^{3}g_im_{T,i}\left[1+(q-1)\frac{m_{T,i}}{T}\right]^{\frac{-q}{(q-1)}}
\end{equation}
 \indent The transverse momentum spectra of primary charged particles measured by the ALICE 
 collaboration~\cite{ALICE} in INEL $p - p$ collisions at $\sqrt{s}$ = 900 GeV $(|\eta| < 0.8)$, normalized 
 to the total number of INEL events, $N_{evt}$, for three different multiplicity 
 selections ($n_{acc}$) together with the Tsallis fits, Eq.~(\ref{tsallisfit}), is shown in Fig. 1.
 
\begin{figure}[ht]
	\begin{minipage}{\columnwidth}
	\centering
	\includegraphics[width=\columnwidth, height = 9.0cm]{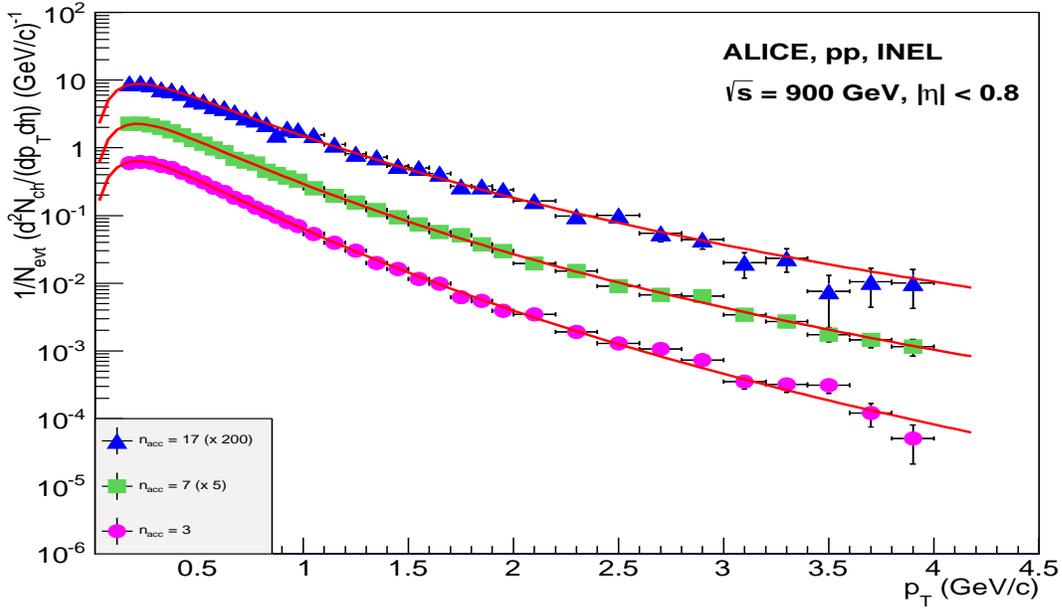}
 \end{minipage}		
\caption{Transverse momentum distributions of charged particles as measured by the 
ALICE collaboration in $p-p$ collisions at $\sqrt{s}$ = 0.9 TeV fitted with Tsallis distribution.}
\label{figALICE}
\end{figure}
 
 \indent The charged hadron yields as measured by the CMS 
 collaboration~\cite{CMS1,CMS2} in the range $|\eta| < $ 2.4 in non-single-diffractive (NSD) events 
 as a function of $p_T$ at all three center-of-mass energies, fitted with Tsallis distribution are 
 shown in Fig. 2. 
 
 \begin{figure}[ht]
\includegraphics[width=\columnwidth, height=9.0cm]{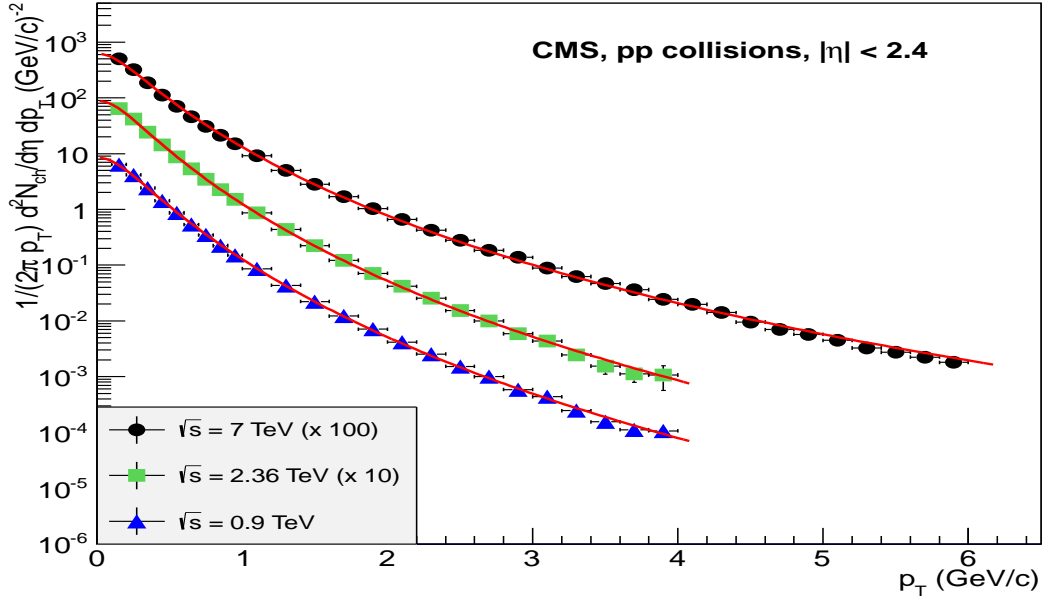}
\caption{\label{figCms} Transverse momentum distributions of charged particles as measured by the 
CMS collaboration in $p-p$ collisions at $\sqrt{s}$ = 0.9, 2.36 and 7  TeV fitted with Tsallis distribution.}
\end{figure}
 
 \indent Fig. 3 shows the charged particle multiplicities as a function of transverse momentum measured by the ATLAS collaboration~\cite{ATLAS} for events with $n_{ch}\ge1$,  $p_T > 500$ MeV and $|\eta| < 2.5$ at $\sqrt{s}$ = 0.9, 2.36 and 7 TeV fitted with Tsallis distribution. \\
\indent The Tsallis fits for the charged particle multiplicities as a function of 
transverse momentum measured by the ATLAS collaboration~\cite{ATLAS} for events 
with $n_{ch}\ge2$,  $p_T > 100$ MeV and $|\eta| < 2.5$ at $\sqrt{s}$ = 0.9 and 7 TeV are shown in Fig. 4.\\
\indent Fig. 5 shows the Tsallis fits on the charged particle multiplicities as a function of 
transverse momentum measured by the ATLAS collaboration~\cite{ATLAS} for events with 
$n_{ch}\ge6$,  $p_T > 500$ MeV and $|\eta| < 2.5$ at $\sqrt{s}$ = 0.9 and 7 TeV.\\
\indent Fig. 6 shows the Tsallis fits for the charged particle cross section for three different beam 
energies, 0.9, 2.76 and 7 TeV measured recently by the ALICE collaboration~\cite{ALICE2}.\\
\indent In all cases the fits are very good.

\section{Results} 

\noindent The transverse momentum spectrum of the primary charged particles measured by the ALICE, CMS and ATLAS collaborations 
and cross section measured by the ALICE  collaboration~\cite{ALICE,CMS1,CMS2,ATLAS,ALICE2} in $p - p$ collisions at 
LHC energies were fitted using the Tsallis distribution given in Eq.~(\ref{tsallisfit}). \\
\indent Though, the kinematical conditions of the data taking of all the three collaborations were different. 
The Tsallis distribution fits well to the measured $p_T$ spectrum as well as the cross section. \\
\indent The values of the Tsallis parameter, $q$, the q-temperature, $T$, and the radius, $R$,
defined as $R \equiv \left({3V/4\pi}\right)^{1/3} $
obtained from the fits of the $p_T$ spectrum are shown in Figs.  7, 8 and 9, respectively. 
The values of the $q$ parameter shows a clear increase with beam energy. The collaborations show consistent compatible results.
It is to be noted that the parameter $q$ can be determined with a fairly high precision.\\
\indent The results for the q-temperature, $T$, shown in Fig. 8 are compatible and show no clear dependence on the beam energy. The only 
points that are clearly higher are those obtained by the ATLAS collaboration for 
data with a high number of charged 
particles ($n_{ch} \geq 6$).\\
\indent The results for the parameter $R$ again show a clear increase with beam energy, which was not seen
in a previous analysis~\cite{dubna}. Again the results are compatible.

\begin{figure}
\includegraphics[width=\columnwidth, height = 9.0cm]{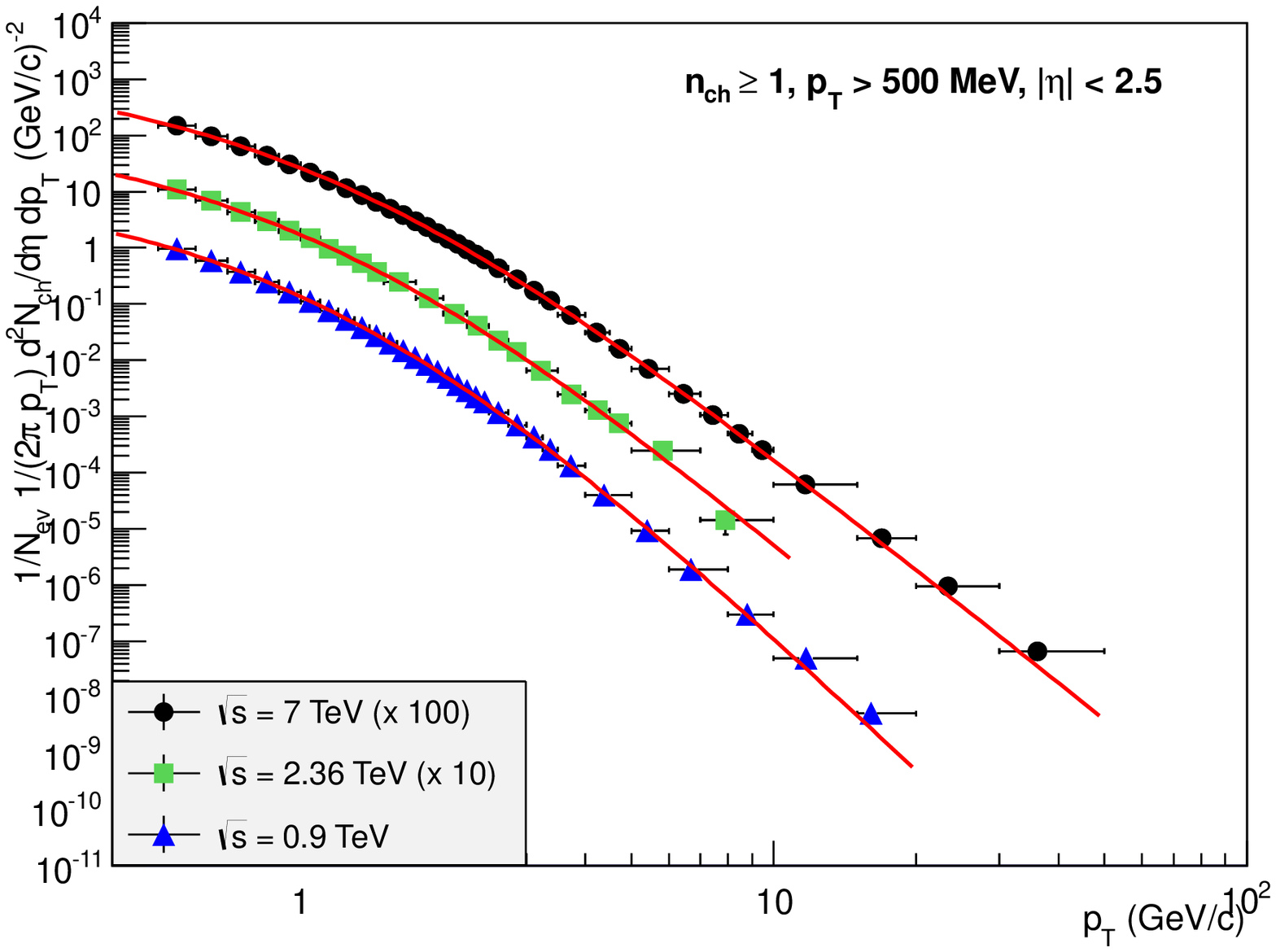}
\caption{\label{figAtlas1} Transverse momentum distributions of charged particles as measured by the ATLAS collaboration at all three center-of-mass energies together with Tsallis fit.}

\includegraphics[width=\columnwidth, height=9.0cm]{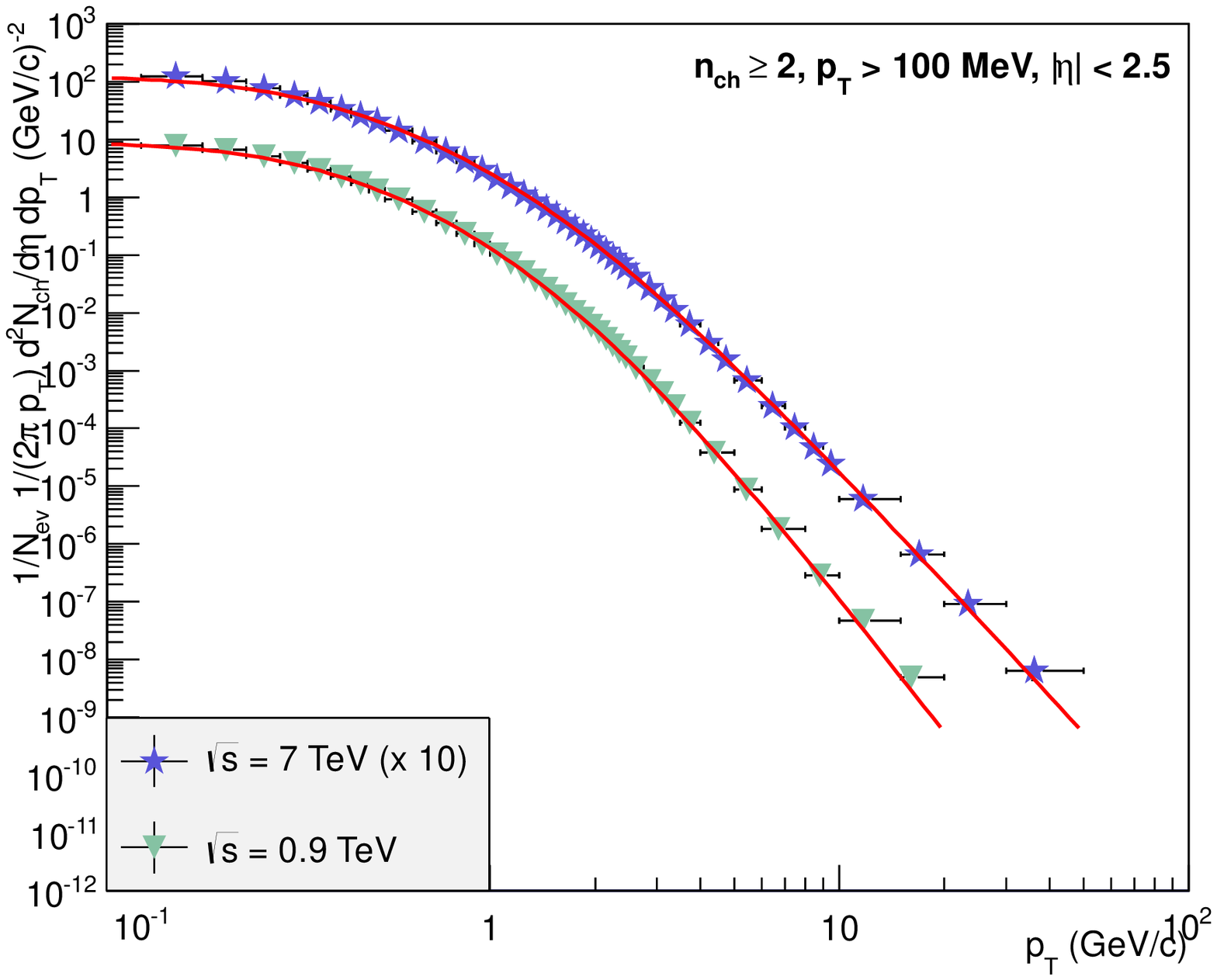}
\caption{\label{figAtlas2} Transverse momentum distributions of charged particles as measured by the ATLAS collaboration in the most inclusive phase-space space region at $\sqrt{s}$ = 0.9 and 7 TeV together with Tsallis fit.}
\end{figure}

\begin{figure}
\includegraphics[width=\columnwidth, height=9.0cm]{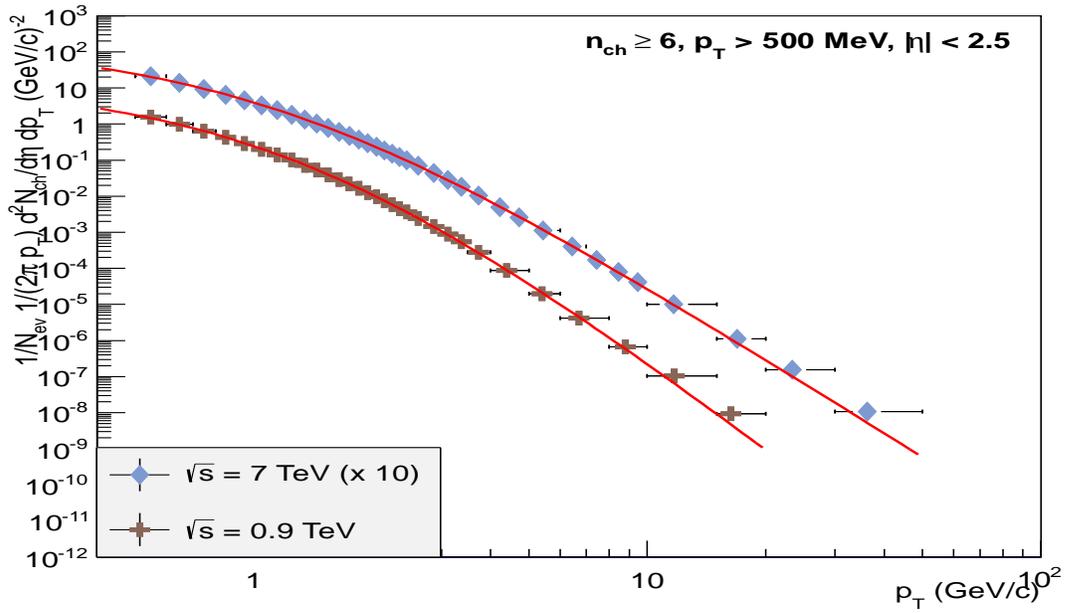}
\caption{\label{figAtlas3} Transverse momentum distributions of charged particles with the smallest contribution from diffractive events as measured by the ATLAS collaboration at $\sqrt{s}$ = 0.9 and 7 TeV together with Tsallis fit.}
\end{figure}
\begin{figure}
\includegraphics[width=\columnwidth, height=9.0cm]{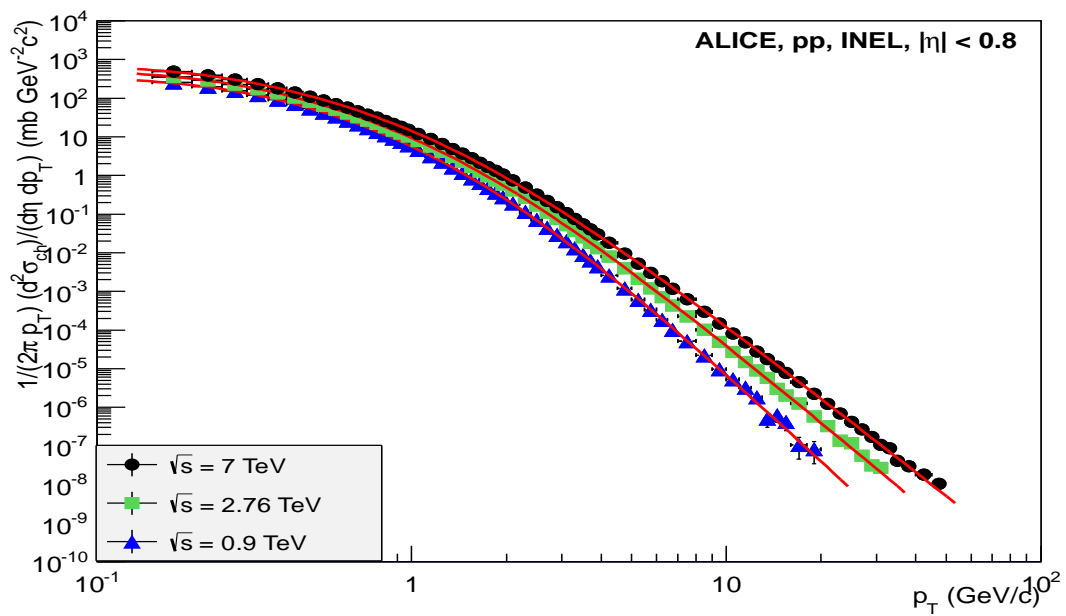}
\caption{\label{figALICE2} Cross section of charged particles as measured
by the ALICE collaboration~\cite{ALICE2} at different energies.}
\end{figure}
\begin{figure}
\includegraphics[width=\columnwidth, height=9.0cm]{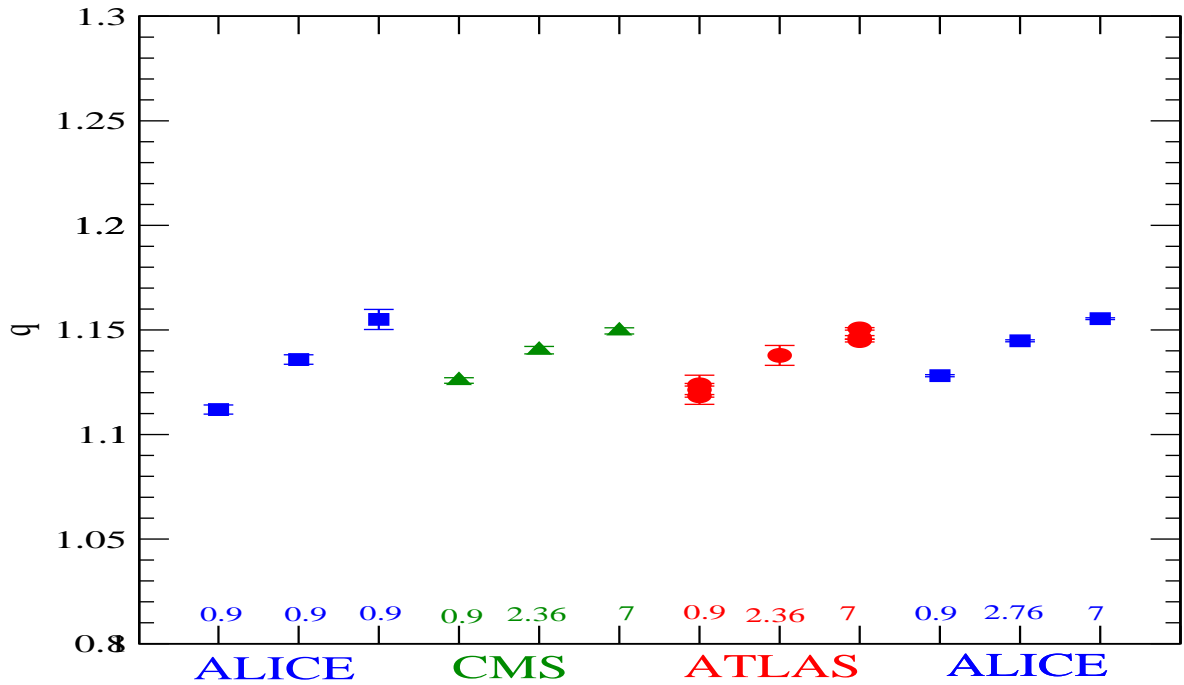}
\caption{\label{figQ} The Tsallis parameter $q$ obtained from  fits to the $p_T$ spectrum. 
	Results from the ALICE collaboration 
	are indicated with square points, CMS by triangles and ATLAS by circles. 
	The beam energy is given on the x-axis. The first three points from ALICE  correspond to 
	different values of accepted charged particles, 
nacc = 3, 7 and 17 respectively. 
}
\end{figure}

\begin{figure}
\includegraphics[width=\columnwidth, height=9.0cm]{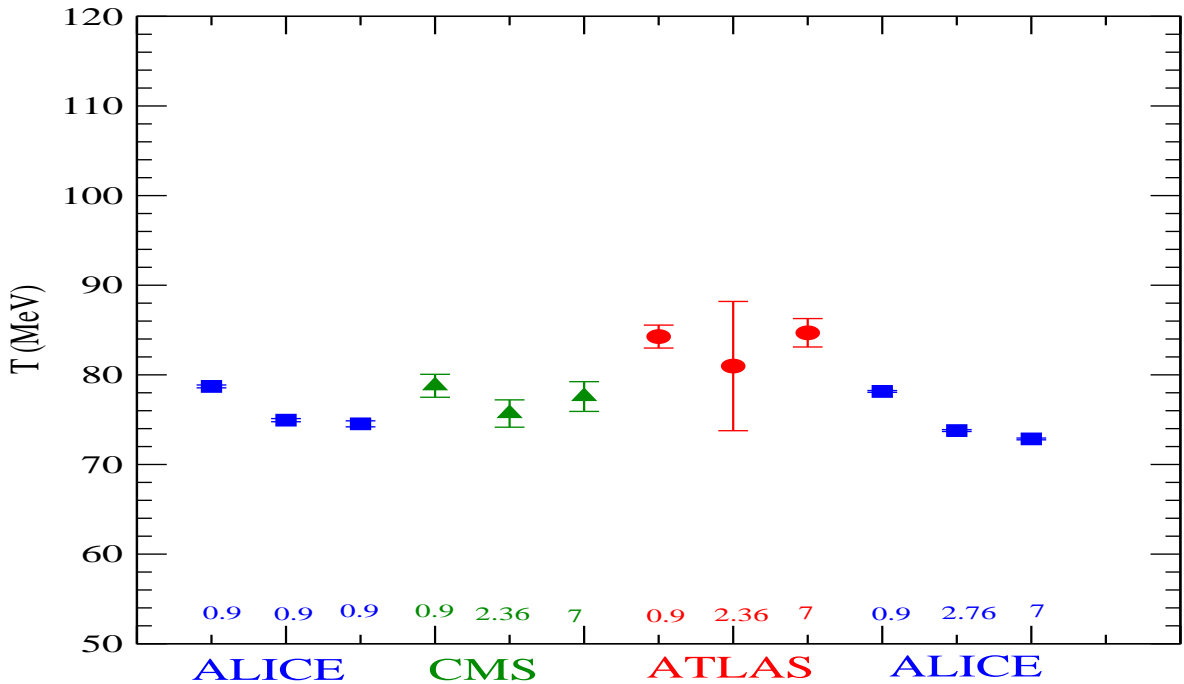}
\caption{\label{figT} The q-temperature, $T$, obtained from  fits to the $p_T$ spectrum. Results from the ALICE collaboration 
	are indicated with square points, from CMS by triangles, those from ATLAS by circles. The first three points were
obtained at a beam energy of 900 GeV and  correspond to different values of accepted charged particles,
nacc = 3, 7 and 17 respectively.}
\end{figure}

\begin{figure}
\includegraphics[width=\columnwidth, height=9.0cm]{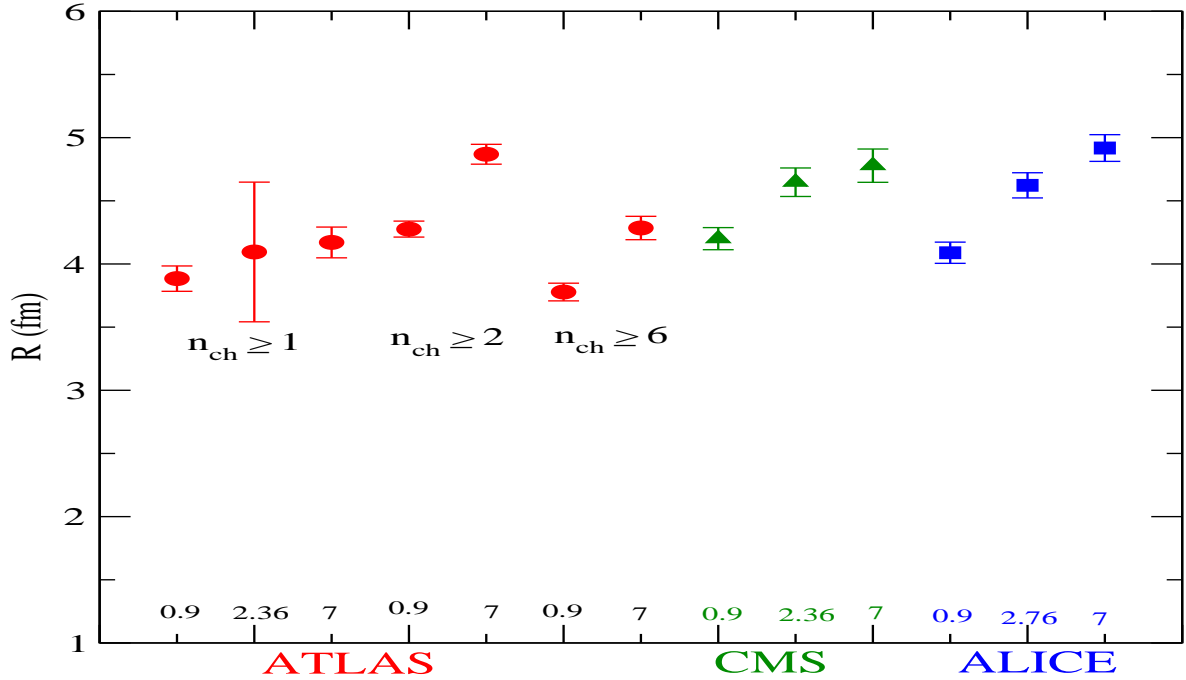}
\caption{\label{figR} The radius, $R$, obtained from  fits to the $p_T$ spectrum.
The results from ATLAS are indicated by circles, CMS by triangles and ALICE by square points. The ATLAS points correspond to 
different acceptances of charged particles. The beam energies are given on the x-axis.}
\end{figure}

\section{Conclusions}
\noindent 
A thermodynamically consistent form of the Tsallis distribution, Eq.~(\ref{tsallisfit}),
gives an excellent description of  the transverse momentum spectra and cross section of the 
primary charged particles measured in $p - p$ collisions at $\sqrt{s}$ = 0.9, 2.36, 7 TeV and 2.76 TeV, respectively.  \\
\indent The values of the Tsallis parameter, $q$, are found between 1.11 to 1.15 and show a clear 
increase with beam energy.\\ 
\indent As observed from the fit of $p_T$ spectra measured by the ALICE, ATLAS and CMS 
collaborations~\cite{ALICE,ATLAS,CMS1,CMS2,ALICE2}, the q-temperature, $T$, is consistent with being constant as a function of beam energy.\\
\indent The values of radius, $R$, shows a small increase with  beam energy.\\
\indent Overall, the values obtained for the Tsallis parameter, $q$, are 
remarkably consistent, a feature which does not become apparent when using the 
parameterization of Eq.~(\ref{BIG}).\\
It is concluded that the hadronic system created in high-energy p - p collisions at central rapidity 
can be seen as obeying Tsallis thermodynamics.\\[0.5cm]
\ack{
We acknowledge useful discussions with C. Tsallis and M. Floris.\\
}

\end{document}